\documentclass[aps,pre,superscriptaddress,footnoteinbib,showpacs,floatfix,twocolumn]{revtex4-1}
\usepackage{graphicx}   
\usepackage{amssymb}
\usepackage{natbib}

\begin{document}
\title{Anomalously slow transport in single-file diffusion with slow binding kinetics} 
\author{Spencer G. Farrell}	
\author{Andrew D. Rutenberg}	
\email{andrew.rutenberg@dal.ca}
\affiliation{Dept. of Physics and Atmospheric Science, Dalhousie University, Halifax, Nova Scotia, Canada B3H 4R2}

\date{\today}
\begin{abstract} 
We computationally study the effects of binding kinetics to the channel wall,  leading to transient immobility, on the diffusive transport of particles within narrow channels, that exhibit single-file diffusion (SFD).  We find that slow binding kinetics leads to an anomalously slow diffusive transport. Remarkably, the scaled diffusivity $\hat{D}$ characterizing transport exhibits scaling collapse with respect to the occupation fraction $p$ of sites along the channel. We present a simple ``cage-physics'' picture that captures the characteristic occupation fraction $p_{scale}$ and the asymptotic $1/p^2$ behavior for $p/p_{scale} \gtrsim 1$.  We confirm that subdiffusive behavior of tracer particles is controlled by the same $\hat{D}$ as particle transport.
\end{abstract}
\maketitle

When large particles randomly diffuse along narrow channels, so that particles cannot change their order, they undergo single-file diffusion (SFD) \cite{Harris1965, Beijeren1983, Taloni2017}.  In SFD, tracer diffusion of individual particles exhibits sub-diffusion within regions of uniform density. SFD has been beautifully visualized with colloidal particles inside engineered micro-channels \cite{Wei2000}, and has been demonstrated for molecular diffusion inside various nanoporous zeolites \cite{Gupta1995,*Smit2008}. SFD is also expected for molecular diffusion inside carbon nanotubes \cite{Mao1999, *Hummer2001}, for flagellin within bacterial flagella \cite{Stern2013},  for the acetylation enzyme $\alpha$-Tat1 within microtubules \cite{Farrell2015}, and for DNA-binding proteins on DNA \cite{Li2009}. 

SFD behavior can be obtained from a system of non-interacting particles undergoing simple diffusion (SD) by exchanging particle labels as their trajectories cross \cite{Harris1965, Levitt1973}. When this label-exchange equivalence applies, SFD systems retains the same collective transport properties as SD. So, collective transport in non-interacting SFD would be independent of particle density \cite{Kutner1981}.

Nevertheless, particle binding or adsorption to narrow channel walls, resulting in transient immobility of bound particles, affects transport.  For SD, such transient immobility rescales diffusivity, since particles only diffuse while unbound. Transient binding has no further effect on SD provided that binding kinetics are sufficiently fast with respect to diffusive timescales \cite{Wilson1948}. Nevertheless, having both bound and unbound particles locally breaks label-exchange equivalence between SFD and non-interacting SD systems --- since the local trajectories of stationary bound particles and  mobile unbound particles differ.  

The density independence of Fickian transport $D_{Fick}$ has {\em only} been derived using label-exchange equivalence \cite{Kutner1981} --- so density-dependent  $D_{Fick}(p)$ is not ruled out in SFD systems with binding kinetics.  Indeed, transitions between mobile and immobile states affects transport for strongly driven systems such as asymmetric exclusion processes or active transport along molecular tracks \cite{Bressloff2013, *Ciandrini2014}. We are interested in whether slow binding kinetics  affects collective transport properties in SFD systems in the linear (Fickian or hydrodynamic) regime, with only weak density gradients driving transport.  How large is any density-dependent effect?  What timescale determines what ``slow'' binding is?  

We can address these questions by adding binding kinetics to a  symmetric simple exclusion process (SSEP) \cite{Spitzer1970, Arratia1983}, which has long been studied as a SFD model of transport.  Our one-dimensional lattice model with spacing $a$ and single-occupancy to enforce SFD is characterized by a local occupation probability $p$, together with rates for single-particle hopping $k_{hop}$, binding $k_{on}$, and unbinding $k_{off}$.  Bound particles are immobilized to the channel wall and do not hop. The equilibrium association constant is  $K_A \equiv k_{on}/k_{off}$, while $K_D \equiv k_{off}/k_{on}$ is the disassociation constant. In the non-interacting limit of vanishing density ($p \rightarrow 0$) we expect to recover the standard result \cite{Wilson1948, Odde1998} of hopping diffusivity ($D_{hop} \equiv k_{hop} a^2/2$) scaled by the fraction of time that particles are unbound, $1/(1+K_A)$ --- i.e. a non-interacting transport diffusivity 
\begin{equation}
	D_0 = \frac{k_{hop} a^2}{ 2} \frac{1}{1+K_A}. 
	\label{D0} 
\end{equation}
We will consider the dimensionless scaled diffusivity 
\begin{equation} 
	\hat{D} \equiv D/D_0. 
\end{equation} 
Deviations from $\hat{D}=1$ indicate a non-trivial effect on transport due to slow binding in SFD. We expect to recover this non-interacting $\hat{D}=1$ in the dilute limit when $p \rightarrow 0$, in the limit of no binding as $K_A \rightarrow 0$, and when $k_{off}$ and $k_{on}$ are sufficiently large.   

We computationally study diffusive transport on our one dimensional system, with a length $L$  lattice, using a fully stochastic simulation algorithm (SSA) \cite{Gillespie1977} for various values of  $K_A$ and $k_{off}$.  We investigate transport properties using an open system with imposed boundary conditions $p_0=1$ and $p_L=0$ --- where $p_i$ is the average occupation of the $i$-th site.  At $i=0$ we immediately inject a new unbound particle whenever the site is empty, and only allow hopping to $i=1$. At $i=L$ we immediately remove any (unbound) particle that arrives from $i=L-1$. This geometry allows us to directly assess transport properties for a range $p \in [0,1]$. We measure the time-average occupations in steady state along with the flux $\Phi$ (average net rate of particles passing any point of the system), which is both uniform along the system and time-independent in steady-state. 

The transport properties of our SSEP system with binding will be unchanged if we consider holes rather than particles since $p_{hole} = 1-p$. Accordingly, the same $D_{Fick}$ applies to both particle and hole transport. Transport is also unchanged with a ``free particle'' model in which unbound particles can exchange labels at the rate $k_{hop}$. We can also consider a ``free hole'' model in which holes are individually tracked and  exchange with each other and with unbound particles at the rate $k_{hop}$. Both free particle and free hole models will have the same transport $D_{Fick}$ as our SFD system. However individual free particles will exhibit mean-square displacements (MSD) that grows linearly with time as $\langle \Delta x_{free}^2 \rangle = 2 D_{MSD}^{part} t$, with a Gaussian distribution of individual displacements after sufficiently long times \cite{Arratia1983}. Similarly individual free holes would exhibit MSD that grow linearly with time with $D_{MSD}^{hole}$. For $p \rightarrow 1$ we expect these free holes to be non-interacting, with $D_{Fick}=D_{MSD}^{hole}$. Similarly for $p \rightarrow 0$ we expect free particles to be non-interacting with $D_{Fick}=D_{MSD}^{part} = D_0$. 

While we are primarily interested in transport properties, a fascinating property of SFD is that tracer trajectories are sub-diffusive, i.e. 
\begin{equation}
	\langle \Delta x_{tr}^2 \rangle = 2 a \frac{1-p}{p} \sqrt{\frac{D_{tr} t}{\pi}},
	\label{subdiffusion}
\end{equation}
where the left-side is the mean-square displacement of individual tracked particles, $p$ is the occupation fraction for a lattice model, $a$ is the lattice spacing, and $t$ is the elapsed time \cite{Harris1965, Levitt1973, Percus1974, vanBeijeren1983}. This result follows quite generally from the hydrodynamic relaxation of initial fluctuations with $D_{tr} = D_{Fick}$ \cite{Alexander1978}.   This result also follows in the $p \rightarrow 0$ and $p \rightarrow 1$ limits from the Gaussian propagators of free particles or holes respectively with $D_{tr}=D_{MSD}=D_{Fick}$, \cite{Hahn1995, Rodenbeck1998}. 

We will numerically check that $D_{tr}=D_{Fick}$ for selected values of $p$ using closed systems with randomly placed particles. For these measurements, as for transport measurements, we wait until steady-state (equilibrium) conditions have been achieved to start our measurements in order to avoid any transients due to initial conditions \cite{Leibovich2013, Alexander1978, Krapivsky2014}.  

\begin{figure}[t]   
\includegraphics[trim = 4mm 4mm 4mm 4mm, clip, width=0.4\textwidth]{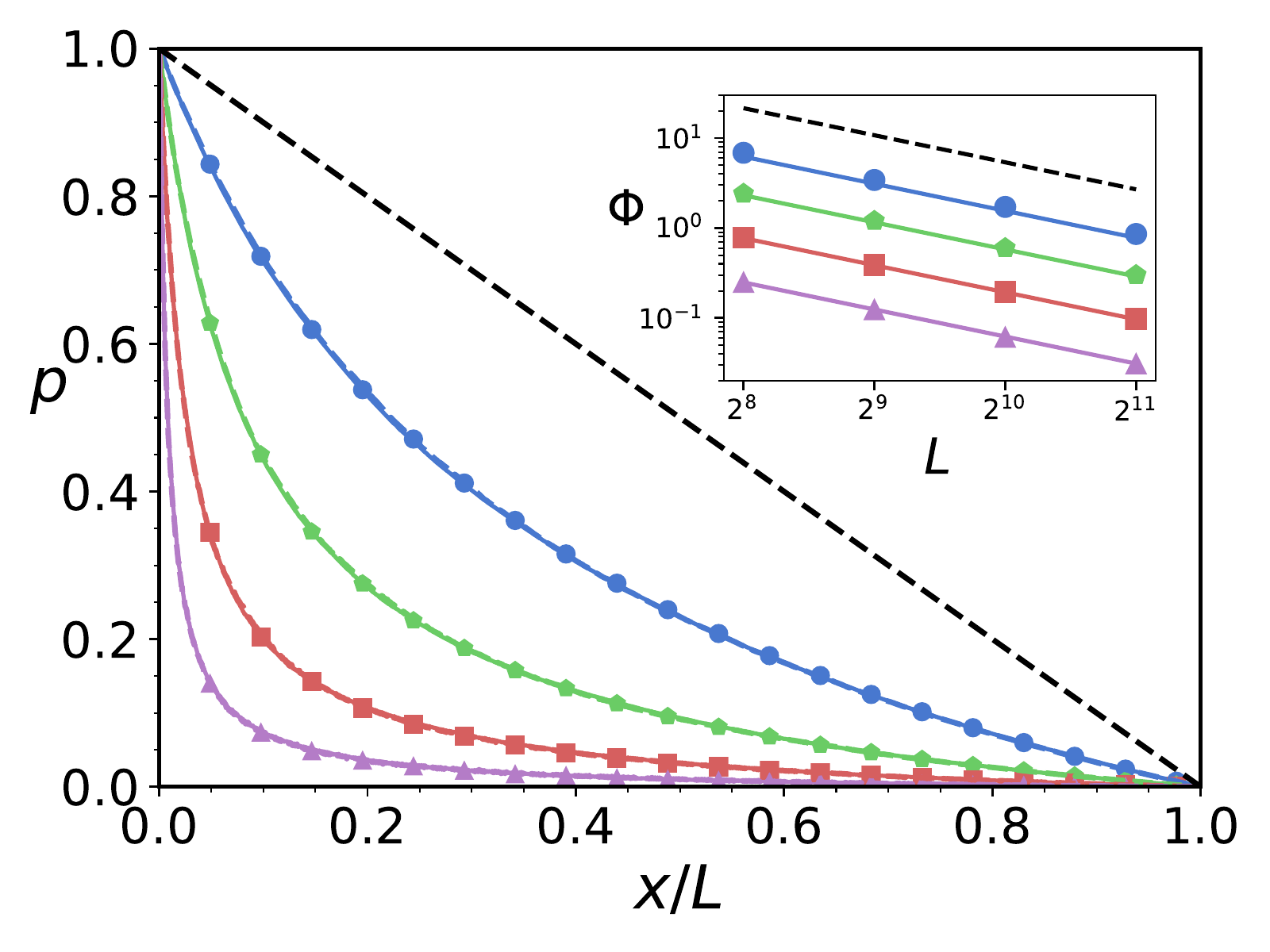} 
\caption{Occupation fraction $p$ vs scaled distance $x/L$, for various $k_{off}$ (solid coloured lines for $L=2048$, superimposed dashed lines with earlier data to demonstrate steady-state,  superimposed points with $L=1024$ to demonstrate linear regime.  All with $K_A=100$, $D_0=2.7 \times 10^5 nm^2/s$ and $a=7nm$.) The diagonal dashed black line is the linear profile expected for concentration-independent transport with $D_{Fick}=D_0$. Instead, stronger concentration dependence is seen with slower $k_{off}$ ($k_{off} = 1/s$, $10/s$, $10^2/s$, $10^3/s$ purple triangles, red squares, green pentagons, and blue circles from bottom to top, respectively). The inset shows the average steady-state flux $\Phi$ ($s^{-1}$) vs $L$, and exhibits the $1/L$ dependence (illustrated by the dashed black and solid coloured lines) expected for the linear ``hydrodynamic'' regime where Fick's law applies.} 
\label{fig:density}
\end{figure}

In Fig.~\ref{fig:density} we show the occupation (concentration) profiles for a selected range of $k_{off}$ values (coloured solid lines). For concentration-independent diffusivities, we would expect a linear profile -- as indicated by the dashed black line. It is apparent that strong concentration-dependent transport is observed,  depending on $k_{off}$.  Slower values of $k_{off}$ have stronger concentration dependence. Our system size $L=2048$ is large enough that our profiles are independent of $L$ at large system sizes --- as illustrated by superimposing (colored points) data with $L=1024$. We have ignored the earliest half of our data to avoid initial transients. Steady-state is demonstrated by superimposing the later half of the ignored data (coloured dashed lines).    Parameters used here, with $K_A=100$, $D_0=2.7 \times 10^5 nm^2/s$ and $a=7nm$, correspond with what we would expect for the $\alpha$-Tat1 acetylation enzyme within the microtubule lumen \cite{Szyk2014, Farrell2015}.

Fick's law says the flux is proportional to the gradient of the density, for sufficiently small gradients:	$D_{Fick} = -\Phi a /\nabla p$, where the local density is $p/a$.  The dashed black line in the inset of Fig.~\ref{fig:density} is the flux expected for transport with $D_{Fick} = D_0$; while our measured flux $\Phi$ has the same $1/L$ scaling (as indicated by the solid coloured lines).  It is apparent that the flux is significantly suppressed due to slow binding. 

\begin{figure}[h]   
\includegraphics[trim = 4mm 4mm 4mm 4mm, clip, width=0.4\textwidth]{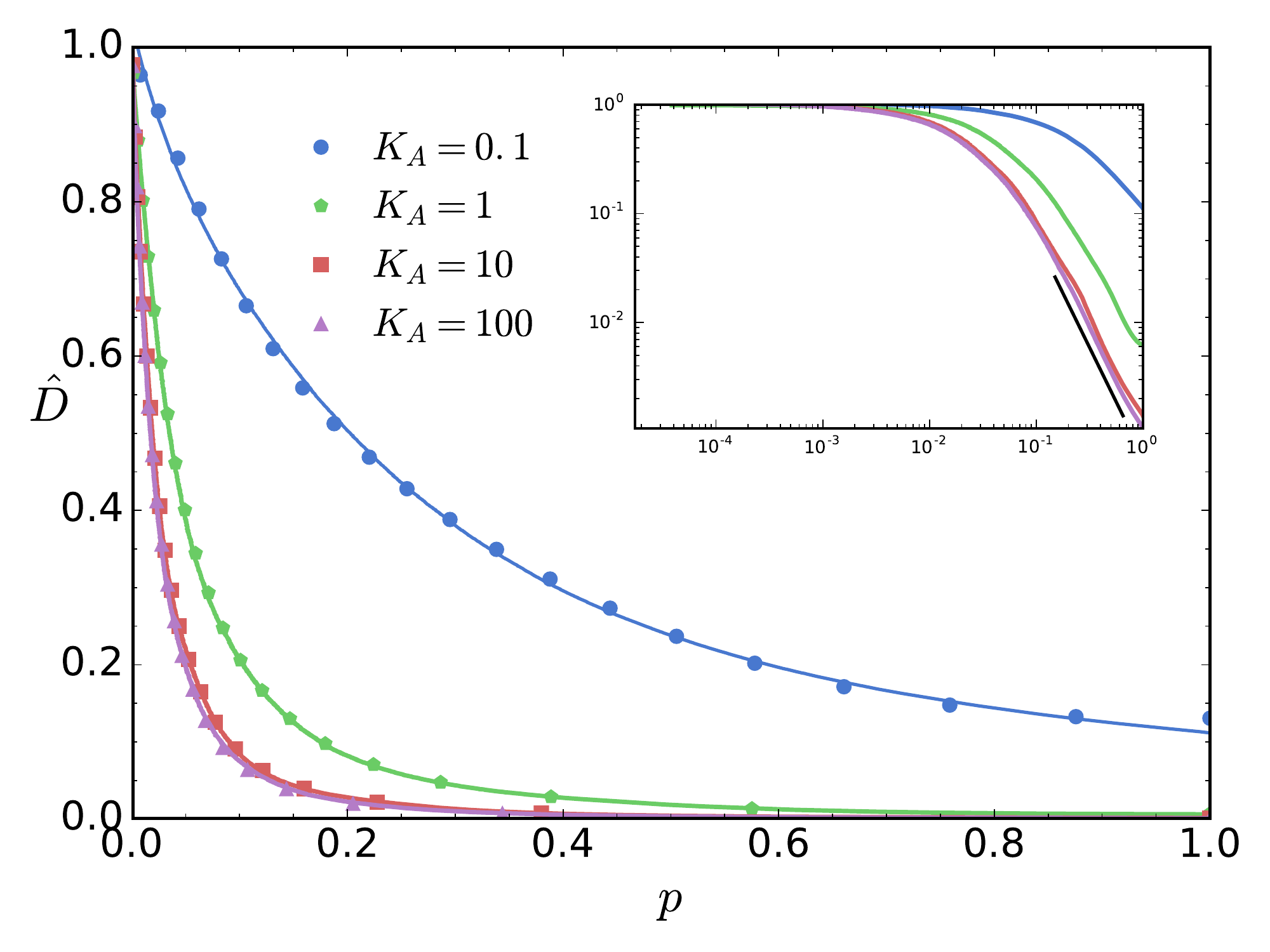}
\caption{Scaled transport $\hat{D} \equiv D_{Fick}(p)/D_0$ vs occupation fraction $p$. At small $p$ the $\hat{D}=1$ limit of SD is recovered, while anomalous slowing is seen for $p>0$. Data is from $L=2048$ (solid lines) and $L=1024$ (points) with $k_{off}=10/s$ but with various $K_A$ as indicated by the legend.  The inset shows the same data on a log-log scale, and emphasizes a characteristic $1/p^2$ dependence at larger $p$ with a solid black line.} 
\label{fig:mobility}
\end{figure}

We can use the measured fluxes and profiles to obtain diffusivities using Fick's law. We use spline-smoothing \footnote{Using splrep in the scipy.interpolate package of Python 2.7.11.} to extract the density gradient from the concentration profiles. In Fig.~\ref{fig:mobility} we show the scaled diffusivities $\hat{D} \equiv D_{Fick}(p)/D_0$ vs. the density $p$. Here, we emphasize the $K_A$ dependence (all with $k_{off}=10/s$). We see that diffusivities decrease monotonically with density, decreasing from the SD result $\hat{D}=1$ at $p=0$ to a maximal suppression at $p=1$. At smaller values of $K_A$, when few particles are bound, we are closer to the SD result at all values of $p$. At larger values of $K_A$ we see a stronger suppression of $\hat{D}$ at larger $p$. In the inset, we highlight a characteristic $1/p^2$ dependence exhibited at larger values of $K_A$ and $p$ with a solid black line on the log-log plot of the same data.  

To explain the physics behind the anomalously slow SFD transport, it is useful to consider a free-particle (or free-hole) model and to focus on the asymptotic $1/p^2$ behavior at large $K_A$ and larger $p$. Particles will be effectively immobilized in cages of size $\ell \approx 2 a/p$ formed by their bound neighbours. The mobile fraction ($1/(1+K_A)$) will escape the cages when they open in a timescale $\tau_{escape} \approx k_{off}^{-1}$. This gives an effective diffusivity $D \approx \ell^2/\tau = 4 a^2 k_{off}/p^2 \sim 1/p^2$ for the mobile fraction, leading to the $1/p^2$ scaling for $\hat{D}$. We expect this to describe, at least approximately, the collective transport $D_{Fick}$. 

For smaller values of $K_A$, the size of the cage is larger since not all particles are bound so $\ell \approx 2 a (1+K_D)/p$. This cage should dominate when the exploration time $\tau_{explore} \approx \ell^2/D_0$ of  particles is less than the escape time $\tau_{escape}$. Setting $\tau_{explore} \approx \tau_{escape}$ determines a characteristic density $p_{scale}$: 
\begin{equation}
	p_{scale} \equiv \sqrt{\frac{k_{on}+k_{off}}{k_{hop}} }\, \left(1 + K_D \right).
	\label{pscale}
\end{equation}

\begin{figure}[b]   
\includegraphics[trim = 4mm 4mm 4mm 4mm, clip, width=0.45\textwidth]{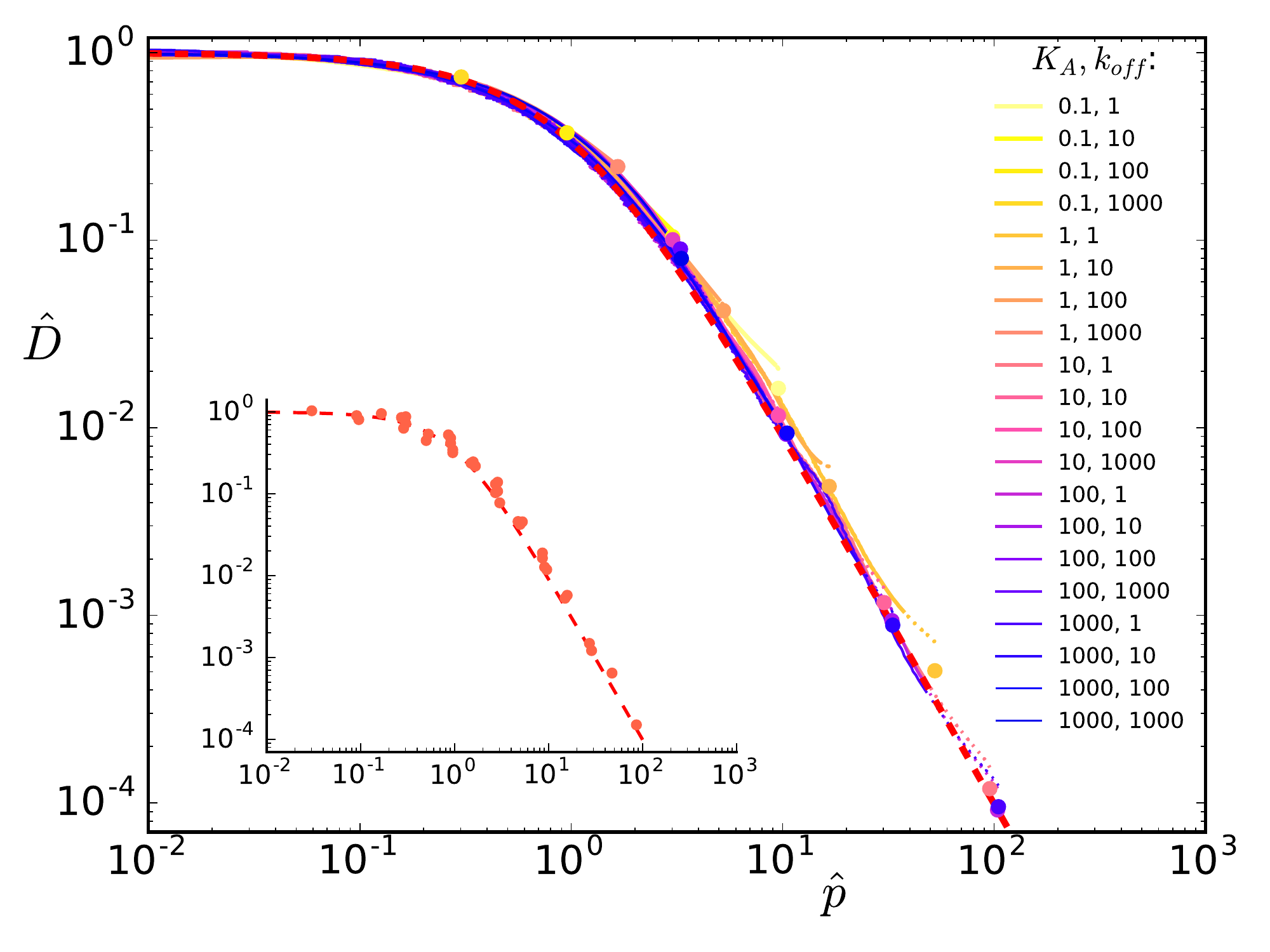}
\caption{Scaled transport $\hat{D} \equiv D/D_0$ vs scaled occupation fraction $\hat{p} \equiv p/p_{scale}$. Lines indicate $D_{Fick}$ as analyzed for various $p$ in an open system of length $L=2048$, and for a range of $K_A$ and $k_{off}$ values indicated by the legend. The lines are dotted for $p>0.9$, where sharp gradients (see Fig.~\ref{fig:density}) begin to affect spline-smoothing results. The coloured points are for single holes  $D_{SFD}^{hole}/D_0$ vs $\hat{p}$ (in a periodic system of $L=2048$).  The inset (red points) indicates tracer-subdiffusion $D_{tr}/D_0$ vs $\hat{p}$ as directly measured for periodic $L=8192$ systems at various average $p$ ($0.1$, $0.3$, and $0.9$), $K_A$ ($0.1$, $1$, and $10$) and $k_{off}$ ($1$, $10$, $100$, $1000$). The dashed red lines are the approximate scaling function $\hat{D}_{scale} \equiv 1/(1+\hat{p}+\hat{p}^2)$. } 
\label{fig:scaledmobility}
\end{figure}

Remarkably, the scaled transport coefficient $\hat{D}$ collapses for all values of $p$ when plotted against the scaled occupation-fraction $\hat{p} \equiv p/p_{scale}$, as shown in Fig.~\ref{fig:scaledmobility}.  As indicated by the legend, this collapse holds over at least a four-decade range of $K_A$, together with three-decades of $k_{off}$ and four-decades of $\hat{p}$. For $\hat{p} \ll 1$ we recover the independent particle limit $\hat{D}=1$. For $\hat{p} \gtrsim 1$ we crossover into the $1/p^2$ scaling ``cage-physics'' regime described above.  The excellent scaling collapse for all $\hat{p}$ indicates that $p_{scale}$ controls the physics also for $\hat{p} \lesssim 1$.

By studying the random walks of isolated (non-interacting) holes we can check our expectation that $D_{MSD}^{hole}=D_{Fick}=D_{tr}$ as $p \rightarrow 1$. With various $p_{scale}$, we can still consider a range of scaled $\hat{p}$. We have studied the mean-square displacement of isolated holes in a (periodically closed) SSE system, extracted $D_{MSD}^{hole} = \langle dx^2 \rangle /(2 t_{max})$ at the latest elapsed time $t_{max}$ \cite{Vestergaard2014} and indicated the results in Fig.~\ref{fig:scaledmobility} with filled circles.  The collapse with $D_{Fick}$ data is excellent. $D_{MSD}^{hole}$ also directly determines tracer particle subdiffusion as $p \rightarrow 1$. Following Eqn.~\ref{subdiffusion}, with $p=(L-1)/L$, we obtain $\langle \Delta x^2_{tr} \rangle = 2a/L (D_{tr} t/\pi)^{1/2}$.  Since moving an isolated hole by $\Delta x$ shifts all intervening particles by one site,  we have exactly that $\langle \Delta x^2_{tr} \rangle = a \langle |\Delta x_{hole}| \rangle/L$.  For Gaussian hole displacements, we have $\langle |\Delta x_{hole}| \rangle = (4 D_{MSD}^{hole} t/\pi)^{1/2}$. This confirms that $D_{Fick}=D_{MSD}^{hole}=D_{tr}$ as $p \rightarrow 1$.

In the limit of $K_A \gg 1$ and $k_{hop} \gg k_{on}$, so that $\hat{p} \gg 1$, an isolated  hole (for $p \rightarrow 1$) can only move by one of the two adjacent bound particles unbinding.  The hole and particle randomly exchange positions before the particle rebinds, leading to $D_{hole} = k_{off} a^2/2$.  Assuming our scaling collapse with $\hat{p}$, this determines the exact amplitude $\hat{D} \simeq 1/\hat{p}^2$ as $\hat{p} \gg 1$. In Fig.~\ref{fig:scaledmobility}, we have indicated the approximate scaling function 
\begin{equation}
	\hat{D}_{scale} \equiv 1/(1+\hat{p}+\hat{p}^2),
	\label{Dscaling}
\end{equation}	 
with a dashed red line, which includes the asymptotic behavior for both small and large $\hat{p}$ together with an empirical correction for intermediate $\hat{p}$.  The agreement is excellent. We can use $\hat{D}_{scale}$ to approximate the constant steady-state flux between arbitrary boundary conditions, $\Phi_{scale} = 2 D_0 p_{scale}/(\sqrt{3}a^2 L) \tan^{-1}((1+ 2 \hat{p})/\sqrt{3})|^{\hat{p}_{high}}_{\hat{p}_{low}}$. The solid coloured lines in the inset of Fig.~\ref{fig:density} are given by $\Phi_{scale}$ with no fitting.

To see if the tracer subdiffusion $D_{tr}$ from Eqn.~\ref{subdiffusion} collapses for general $p$, we have measured $\langle \Delta x_{tr}^2 \rangle$ for all particles and extracted $D_{tr}$ for $p=0.1$, $0.3$, and $0.9$ each for a range of values of both $K_A$ ($0.1$, $1$, and $10$) and $k_{off}$ ($1$, $10$, $100$, and $1000$). These were measured in periodic length $L=8192$ systems, and $D_{tr}$ was extracted by a fit to Eqn.~\ref{subdiffusion}. The results are shown in the inset of Fig.~\ref{fig:scaledmobility} together with the approximate scaling function as a dashed red line. We see that $D_{tr}$ collapses on the same curve as $D_{Fick}$ for various $p$, confirming our expectation that $D_{Fick} = D_{tr}$. 


In summary, we have explored the effect of transient particle binding and immobility in single-file diffusion (SFD). We have found a strong density dependent diffusivity $D(p)$ that describes both particle transport and the subdiffusion of tracked particles. At low densities, or with fast binding kinetics, we recover the standard $D_0$ (Eqn.~\ref{D0}), but at higher densities or with faster kinetics we observe anomalous slowing with $\hat{D} \equiv D/D_0 \sim 1/p^2$. 

We have observed a collapse of $\hat{D}$ when the density $p$ is scaled by $p_{scale}$ (Eqns.~\ref{pscale} and \ref{Dscaling}) --- this scaling is over $4$ decades in $K_A$, $3$ decades in $k_{off}$, and over the entire range of densities $p \in [0,1]$. We propose that the physics that describes $p_{scale}$ is due to transient trapping of mobile particles between cages formed by bound particles. Mobile particles undergo random walks with step size characterized by the cage size, and step time characterized by the unbinding rate. This is qualitatively similar to the $1d$ escape of particles from regions with switching boundaries described by Holcman {\em et al} \cite{Reingruber2009, Holcman2013}. However this analogy is not exact, since our cages are dynamical while Holcman {\em et al} have escape from a single fixed cage. 

We have demonstrated anomalous transport directly with $D_{Fick}$, but we expect on general terms that this also controls tracer subdiffusion with $D_{tr} = D_{Fick}$ in Eqn.~\ref{subdiffusion}. We have confirmed this directly (see inset of Fig.~\ref{fig:scaledmobility}). We have also directly checked the free-hole limit $p \rightarrow 1$, where $D_{MSD}^{hole}=D_{tr}=D_{Fick}$. A very general approach to large deviations in SFD systems with a density-dependent $D_{Fick}(p)$ and mobility $\sigma(p)$, together with the expectation that $D_{Fick}=D_{tr}$ \cite{Alexander1978}, allows us to conclude that $\sigma(p) = 2p (1-p)/a D(p)$ and also thereby determines all moments of the current fluctuations \cite{Krapivsky2014, *Bodineau2004, *Bertini2002}. 

There are  indications of strong binding effects within carbon nanotubes (CNT) \cite{Hummer2001, Secchi2016}. Our results indicate that transient binding can have further anomalous effects as the CNT diameter approaches molecular diameters \cite{Hummer2001}.  Earlier studies in biomedical or biophysical systems with SFD effects on transport without binding \cite{Yang2010, Stern2013, Farrell2015} should also be revisited in light of transient binding.   

While any violation of the single-file condition destroys the asymptotic (long-time) subdiffusion of tracer-particles \cite{Lucena2012}, the anomalous transport that we have described will not be destroyed. Rather, we expect that a finite rate of particle crossing will simply renormalize the cage escape rate $k_{off}$ in $p_{scale}$. In other words, small particle crossing rates will moderate but not eliminate the anomalous suppression of transport due to slow binding that we have described. 

We thank ACENET and Compute Canada for computational resources. ADR thanks the Natural Sciences and Engineering Research Council (NSERC) for operating grant RGPIN-2014-06245. SF thanks NSERC for a CGSM fellowship.  

\bibliography{refs}   
\end{document}